\documentclass[aps,prl,showpacs,twocolumn,amssymb,groupedaddress]{revtex4}

\usepackage{times}      
\usepackage{graphicx}   
\usepackage{epsfig}      
\usepackage{dcolumn}    

\begin{document}
\title{Gravitational-Wave Emission from Rotating Gravitational Collapse
in three Dimensions}

\author{L.~Baiotti$^{1}$, I.~Hawke$^{2,4}$, L. Rezzolla$^{1,3}$, and
E. Schnetter$^{2}$}

\affiliation{$^1$SISSA, International School for
        Advanced Studies and INFN, Via Beirut 2, 34014 Trieste, Italy}
\affiliation{$^2$Max-Planck-Institut f\"ur Gravitationsphysik,
        Albert-Einstein-Institut, 14476 Golm, Germany}
\affiliation{$^3$Department of Physics, Louisiana State University, Baton
        Rouge, LA 70803 USA}
\affiliation{$^4$School of Mathematics, University of Southampton,
  Southampton SO17 1BJ, UK}

\date{\today}

\begin{abstract} 
        We present  the first calculation of  gravitational wave emission
        produced  in  the gravitational  collapse  of uniformly  rotating
        neutron   stars  to  black   holes  in   fully  three-dimensional
        simulations.   The   initial  stellar  models   are  relativistic
        polytropes  which  are  dynamically  unstable  and  with  angular
        velocities ranging from slow rotation to the mass-shedding limit.
        An  essential   aspect  of  these  simulations  is   the  use  of
        progressive  mesh-refinement techniques which  allow to  move the
        outer  boundaries of  the computational  domain to  regions where
        gravitational  radiation   attains  its  asymptotic   form.   The
        waveforms have been extracted using a gauge-invariant approach in
        which the  numerical spacetime is matched  with the non-spherical
        perturbations of a Schwarzschild spacetime.  Overall, the results
        indicate  that  the  waveforms   have  features  related  to  the
        properties  of the  initial  stellar models  (in  terms of  their
        $w$-mode oscillations)  and of the newly  produced rotating black
        holes  (in  terms  of   their  quasi-normal  modes).   While  our
        waveforms are  in good qualitative agreement  with those computed
        by  Stark and  Piran in  two-dimensional simulations~\cite{sp85},
        our amplitudes are about one  order of magnitude smaller and this
        difference  is  mostly  likely   due to our  less  severe  pressure
        reduction.    For  a   neutron  star   rotating   uniformly  near
        mass-shedding and collapsing at 10 kpc, the signal-to-noise ratio
        computed uniquely  from the  burst is $S/N  \sim 0.25$,  but this
        grows to be $S/N \lesssim 4$ in the case of LIGO II.
\end{abstract}



\maketitle


        The study of the gravitational collapse of rotating stars to
black holes is a cornerstone of any theory of gravity and a long standing
problem in general relativity.  Over the years, numerous approaches have
been applied and several different techniques developed to tackle this
problem which is not just academic. Indeed, important issues in
relativistic astrophysics awaiting clarification, such as the mechanism
responsible for $\gamma$-ray bursts, may be unveiled with a more detailed
understanding of the physics of gravitational collapse in rotating and
magnetized stars. Furthermore, the study of gravitational collapse will
provide the waveforms and the energetics of one of the most important
sources of gravitational radiation.

        In the absence of symmetries, no analytic solutions are known
that describe the gravitational collapse to a black hole and it is
therefore through numerical relativity simulations that one can hope to
improve our knowledge of this process under realistic
conditions. However, this is not an easy task and the modelling of black
hole spacetimes with collapsing matter-sources in multidimensions is one
of the most formidable efforts of numerical relativity. This is due both
to the inherent difficulties and complexities of the system of equations
which is to be solved ({\it i.e.}\ the Einstein equations coupled to the
relativistic hydrodynamics equations) and to the vast computational
resources needed in three-dimensional (3D) evolutions.

        In addition to the technical difficulties due to the accurate
treatment of the hydrodynamics involved in the collapse, the precise
calculation of the gravitational radiation emitted in the process is
particularly challenging as the energy released in gravitational waves is
much smaller than the total rest-mass energy of the system. Indications
of the difficulties inherent to the problem of calculating the
gravitational-wave emission in rotating gravitational collapse have
emerged in the first and only work on this, which dates back to almost 20
years ago~\cite{sp85}. In 1985, in fact, in a landmark work in numerical
relativity, Stark and Piran used their axisymmetric general-relativistic
code to evolve rotating configurations and to compute the gravitational
radiation produced by their collapse to black holes. The results referred
to initial configurations consisting of polytropic stars which underwent
collapse after the pressure was reduced by a factor ranging from 60\% up
to 99\% for the rapidly rotating models. The initial data effectively
consisted of spherically-symmetric solutions with a uniform rotation
simply ``added'' on. Not being stationary solutions of the Einstein
equations, these stars could reach dimensionless spins up to $a\equiv
J/M^2=0.94$, with $J$ and $M$ the angular momentum and mass of the star,
respectively.

        Overall, their investigation revealed that while the nature of
the collapse depended on the parameter $a$, the form of the waves
remained roughly the same over the entire range of the values of $a$,
with the amplitude increasing with $a$. Particularly important was the
evidence that, despite the complex dynamics of the matter during the
collapse, the gravitational-wave emission could essentially be related to
the oscillations of a perturbed black hole spacetime.

        In recent years many studies have extended to three spatial
dimensions the investigation of gravitational collapse to black
holes~\cite{Shibata99e,Shibata03,Duez04}. Despite the improvements
in the evolution of the hydrodynamics and of the spacetime achieved by
these simulations, none of them has addressed the problem of the
gravitational-wave emission. The reason for this is to be found in the
small amplitude of the signal, often below the truncation error of the 3D
simulations, but most importantly in the fact that all of the above
calculations made use of Cartesian grids with uniform spacing.  With the
computational resources currently available, this choice places the outer
boundaries too close to the source to detect gravitational radiation.

        Using a recently developed code for the solution of the Einstein
equations in non-vacuum spacetimes, the {\tt Whisky} code, we have
investigated the collapse of rapidly rotating relativistic stars to Kerr
black holes~\cite{betal04}. An important aspect of these simulations is
the detailed study of the geometrical and dynamical properties of both
the {\it apparent} and {\it event} horizons, allowing for the
determination of both the mass and angular momentum of the black hole
with an accuracy unprecedented for a 3D simulation. In turn, these
measures have set upper limits on the energy and angular momentum lost
during the collapse in the form of gravitational radiation, the first
such estimates coming from 3D calculations.

        However, as in previous works ({\it e.g.}\
refs.~\cite{Shibata99e,Shibata03,Duez04}), the simulations reported
in~\cite{betal04} made use of numerical grids with uniform spacing and
thus with outer boundaries very close to the initial position of the
stellar surface. With these restrictions, the gravitational radiation
extracted does not provide interesting information besides the obvious
change in the quadrupole moment of the background spacetime. As we will
show below, the use of progressive mesh-refinement (PMR) techniques has
removed these restrictions, enabling us to place the outer boundaries of
the computational domain at very large distances from the collapsing
star. This conceptually simple but technologically challenging
improvement has two important physical consequences. Firstly, it reduces
the influence of inaccurate boundary conditions at the outer boundaries
of the domain whilst retaining the required accuracy in the region where
the black hole forms. Secondly, it allows the wave-zone to be included in
the computational domain and thus to extract the gravitational waves
produced in the collapse.

\begin{figure*}
\centerline{
\psfig{file=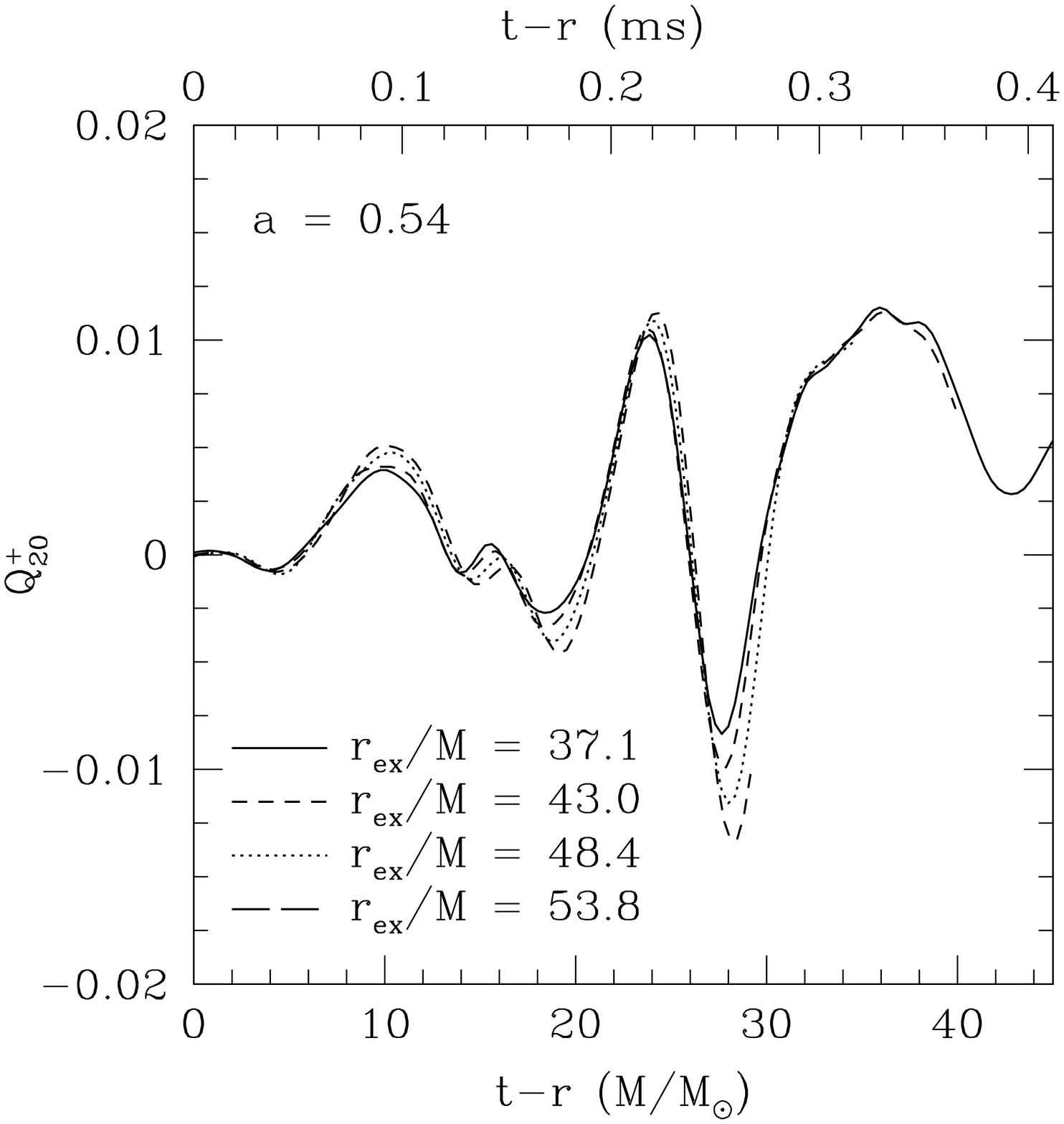,angle=0,width=7.5cm}
\hspace{1.cm}
\psfig{file=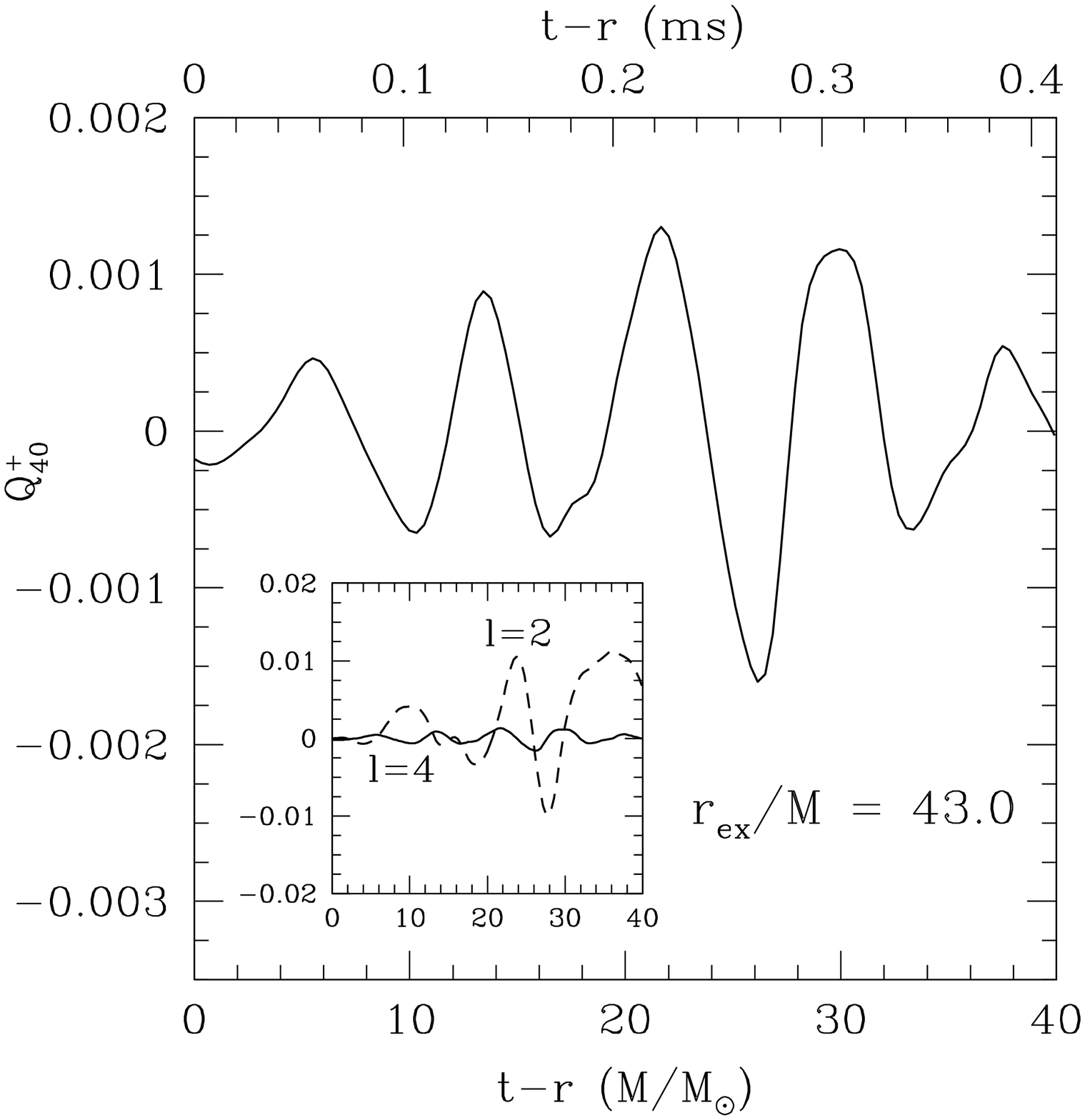,angle=0,width=7.5cm}
        }
\vspace{-0.25cm}
\caption{The left panel shows the $\ell=2$, even-parity perturbation as
extracted by observers at different positions $r_{\rm ex}$ expressed in
retarded time. The right panel refers to the $\ell=4$ mode, with the
inset offering a comparison in the amplitudes of the two modes.}
\vspace{-0.5cm}
\label{waveforms_at_70}
\end{figure*}

        In practice, we have adopted a Berger-Oliger prescription for the
refinement of meshes on different levels~\cite{Berger84} and used the
numerical infrastructure described in~\cite{Schnetter-etal-03b}. In
addition to this, we have also implemented a simplified form of
adaptivity in which new refined levels are added at predefined positions
during the evolution.  More specifically, given an initial stellar model
of mass $M$ and equatorial coordinate radius $R$, our initial grid
consists of four levels of refinement, with the innermost one covering
the star with a typical resolution of $\Delta x \sim 0.17\,M$ and with
the outermost having a typical resolution of $\Delta x \sim 1.38\,M$ and
extending up to $\sim 82.5\,M \simeq 20.9\ R$. As the collapse proceeds
and the star occupies smaller portions of the computational domain, three
more refined levels are added one by one, nested in the four original
ones.  By the time the simulation is terminated at $\sim 81.5\,M$, the
finest typical spatial resolution is $\Delta x \sim 0.02\,M$. A detailed
discussion of the grid and of its evolution will be given
in~\cite{betal05}.

        The initial data for our simulations is the same described in
\cite{betal04} and basically consists of axisymmetric rotating
relativistic stars, calculated as equilibrium solutions of the Einstein
equations in a compactified domain and in polar coordinates. For a direct
comparison with the results in~\cite{sp85} and because no shock is
observed during the collapse, the stars are modelled with a polytropic
equation of state (EOS) $p=K \rho^{\Gamma}$, with $\Gamma=2$ and with the
polytropic constant which is initially $K_{_{\rm ID}}=100$. Once
secularly unstable solutions are found along sequences of fixed angular
momentum or fixed rest-mass, we consider initial models that have the
same axis ratios but slightly larger central energy densities and are
dynamically unstable.  Hereafter we will restrict the discussion to the
collapse of the most rapidly rotating dynamically unstable model, namely
model D4 in~\cite{betal04}, which represents a star of mass
$M=1.861\,M_{\odot}$, circumferential equatorial radius $R_e=14.25$ km
and rotating close to the mass-shedding limit with $a=0.54$ ({\it cf.}
Table I of~\cite{betal04}). The discussion of the emission from stellar
models rotating at smaller velocities will be presented
in~\cite{betal05}.

        Although model D4 is dynamically unstable, it is very close to a
stationary solution and we trigger its collapse by reducing the pressure
by $\lesssim 2\%$, with a change in the value of the polytropic
constant. After the perturbation is introduced, the constraint equations
are again solved to enforce the constraint violation to be at the
truncation-error level. As a contrast, in ref.~\cite{sp85}, rapid
rotation was added to a spherically symmetric solution and the large
pressure depletion inevitably produced a dependence of the results on the
amount of pressure reduction. Hereafter we will concentrate on the
gravitational-wave emission, but a discussion of the dynamics of matter
and trapped surfaces can be found in~\cite{betal04}.

        While several different methods are possible for the extraction
of the gravitational-radiation content in numerical spacetimes, we have
adopted a gauge-invariant approach in which the spacetime is matched with
the non-spherical perturbations of a Schwarzschild black hole (see
refs.~\cite{acs98,Rupright98,cs99} for applications to Cartesian
coordinates grids). In practice, a set of ``observers'' is placed on
2-spheres of fixed coordinate radius $r_{\rm ex}$, where they extract the
gauge-invariant, odd $Q^{({\rm o})}_{\ell m}$ and even-parity
$\Psi^{({\rm e})}_{\ell m}$ metric perturbations~\cite{moncrief74}. Here
$\ell, m$ are the indices of the angular decomposition and we usually
compute modes up to $\ell=5$ with $m=0$; modes with $m \neq 0$ are
essentially zero because of the high degree of axisymmetry in the
collapse. Validations of this approach in 3D vacuum spacetimes can be
found in refs.~\cite{cs99,Rezzolla99a,betal00}, while its use with matter
sources has first been reported in~\cite{fetal02}.

        Although the position of such observers is arbitrary and the
information they record must be the same for waves extracted in the
wave-zone, we place our observers between $40\,M$ and $50\,M$ from the
centre of the grid so as to maximize the length of the extracted
waveform. We note that while a similar choice was made in
ref.~\cite{sp85}, it still provides only an approximate description of
what would be observed at spatial infinity.

        Using the odd and even-parity perturbations $Q^{\times}_{\ell m}
= \lambda Q^{({\rm o})}_{\ell m}$ and $Q^{+}_{\ell m}= \lambda\Psi^{({\rm
e})}_{\ell m}$, where $\lambda \equiv \sqrt{{2(\ell+2)!} / {(\ell-2)!}}$,
we report in Fig.~\ref{waveforms_at_70} the lowest-order multipoles for
$Q^{+}_{\ell m}$ with the offset produced by the stellar quadrupole
removed~\cite{betal05}. The left panel, in particular, refers to the
$\ell=2$ mode as extracted by four different observers at increasing
distances and expressed in retarded time. The right panel instead refers
to the $\ell=4$ mode, with the inset giving a comparison between the two
modes and showing that the gravitational-wave signal is essentially
quadrupolar, with the $\ell=2$ mode being about an order of magnitude
larger than the $\ell=4$ mode.

        The very good overlap of the waveforms measured at different
positions is important evidence that the extraction has been performed in
the wave-zone, since the invariance under a retarded-time scaling is a
property of the solutions of a wave equation. The overlap disappears if
the outer boundary is too close or when the waves are extracted at
smaller radii. A similar overlap is seen also for the $\ell=4$ mode (not
shown).

\begin{figure}
\centerline{
\psfig{file=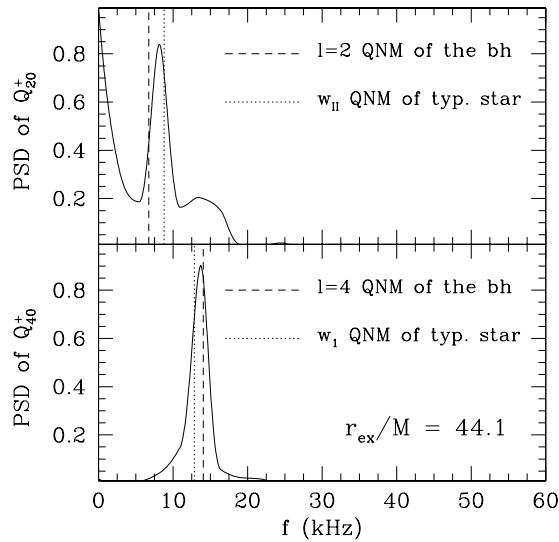,angle=0,width=7.5cm}
        }
\vspace{-0.25cm}
\caption{Power spectra of the waveforms reported in
Fig.~\ref{waveforms_at_70}. The dashed vertical lines indicate the
frequencies of the QNMs of a black hole, while the dotted ones the $w_1$
and $w_{_{\rm II}}$ modes of a typical star.}
\vspace{-0.5cm}
\label{fft_20}
\end{figure}

        Another indication that the waveforms in
Fig.~\ref{waveforms_at_70} are an accurate description of the
gravitational radiation produced by the collapse comes by analysing their
power spectra. The collapse, in fact, can be viewed as the rapid
transition between the spacetime of the initial equilibrium star and the
spacetime of the produced rotating black hole. It is natural to expect,
therefore, that the waveforms produced in this process will reflect the
basic properties of both spacetimes and in particular the fundamental
frequencies of oscillation. We validate this in Fig.~\ref{fft_20}, where
we show the power spectral densities (PSD) of the waveforms of the metric
perturbations $Q^+_{20}$ and $Q^+_{40}$ reported in
Fig.~\ref{waveforms_at_70} (the units on the $y$-axis are arbitrary). The
upper panel of Fig.~\ref{fft_20}, in particular, shows the PSD of
$Q^+_{20}$ and compares it with the frequencies of the $\ell=2, m=0$
quasi-normal mode (QNM) of a Kerr black hole with $M=1.861\,M_{\odot}$
and $a=0.6$~\cite{l85} (dashed line at 6.7 kHz) as well as with the first
$w_{_{\rm II}}$ ``interface'' mode for a typical compact star with
$M=1.27\,M_{\odot}$ and $R=8.86$ km (dotted line at 8.8
kHz)~\cite{ks99}. Similarly, the lower panel of Fig.~\ref{fft_20}, shows
the PSD of $Q^+_{40}$ comparing it with the $\ell=4, m=0$ QNM of a
Schwarzschild black hole~\cite{ks99} (dashed line at 14.0 kHz) and the
first $w_1$ ``curvature'' mode~\cite{ks99} (dotted line at 12.8
kHz)~\cite{note}. It is quite apparent that both peaks in the PSDs are
rather far from the maximum sensitivity area of modern interferometric
and bar detectors.

        Although the waveforms have very short duration with very broad
PSDs, Fig.~\ref{fft_20} shows that these have strong and narrow peaks (a
similar behaviour can be shown to be present also for the odd-parity
modes~\cite{betal05}). Indeed, the excellent agreement between the
position of these peaks and the fundamental frequencies of the vacuum and
non-vacuum spacetimes is an important confirmation of the robustness of
the results obtained.

\begin{figure}
\centerline{
\psfig{file=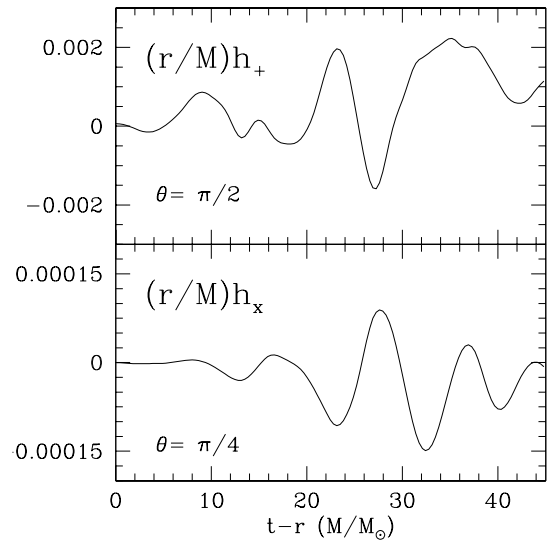,angle=0,width=7.5cm}
        }
\vspace{-0.25cm}
\caption{TT gravitational-wave amplitudes extracted by the detector at
$r_{\rm ex}= 37.1\,M$ and for two different inclination angles.}
\vspace{-0.25cm}
\label{hp_hc}
\end{figure}

        Using the extracted gauge-invariant quantities it is also
possible to calculate the transverse traceless (TT) gravitational-wave
amplitudes in the two polarizations $h_+$ and $h_{\times}$ as
\begin{equation}
h_{+}-{\mathrm i}h_{\times}=\frac{1}{2r}\sum_{\ell, m}
        \left(Q^{+}_{\ell m}-{\mathrm i}
        \int_{-\infty}^{t}\!\!\!\!\!\!
        Q^{\times}_{\ell m}(t')dt'
        \right)\,_{_{-2}}Y^{\ell m}, 
\end{equation}
where $_{_{-2}}Y^{\ell m}$ is the $s=-2$ spin-weighted spherical
harmonic. Because of the small amplitude of higher-order modes, the TT
wave amplitudes can be simply expressed as \hbox{$h_{+} \simeq
h_{+}(Q^{+}_{20},Q^{+}_{40})$} and $h_{\times} \simeq
h_{\times}(Q^{\times}_{30},Q^{\times}_{50})$, where $Q^{\times}_{30} \gg
Q^{\times}_{50}$. Their waveforms are shown in Fig.~\ref{hp_hc} for the
detector at $r_{\rm ex}= 37.1\,M$ and for two different inclination
angles~\cite{betal05}. Note that the amplitudes in the cross polarization
are about one order of magnitude smaller than those in the plus
polarization, with the maximum amplitudes in a ratio $|(r/M)
h_{\times}|_{\rm max}/|(r/M) h_{+}|_{\rm max} \simeq 0.06$. Here, the
odd-parity perturbations, which are zero in spacetime with axial and
equatorial symmetries, are just the result of the coupling, induced by
the rotation, with the even-parity perturbations.

        A precise comparison of the amplitudes in Fig.~\ref{hp_hc} with
the corresponding ones calculated in~\cite{sp85} is made difficult by the
differences in the choice of initial data and, in particular, by the
impossibility of reaching $a \gtrsim 0.54$ when modelling consistently
stationary polytropes in uniform rotation. However, when interpolating
the results in~\cite{sp85} for the relevant values of $a$, we find a very
good agreement in the form of the waves, but also that our estimates are
about one order of magnitude smaller, with $|(r/M) h_{+}|_{\rm max}
\simeq 0.00225$. Furthermore, we observe the amplitude of the
gravitational waves to increase with the pressure reduction, thus
suggesting that the origin of the difference is related mainly to
this~\cite{betal05}.

        Following ref.~\cite{t87} and considering the optimal sensitivity
of VIRGO for the burst signal only, we set an upper limit for the
characteristic amplitude produced in the collapse of a rapidly and
uniformly rotating polytropic star at 10 kpc to be $h_c = 5.77 \times
10^{-22}(M/M_{\odot})$ at a characteristic frequency {$f_c=931$ Hz.} In
the case of LIGO I, instead, we obtain $h_c = 5.46 \times
10^{-22}(M/M_{\odot})$ at {$f_c=531$ Hz}.  In both cases, the
signal-to-noise ratio is $S/N \sim 0.25$, but this can grow to be
$\lesssim 4$ in the case of LIGO II. These ratios could be increased
considerably with the detection of the black hole ringing following the
initial burst~\cite{betal05}. Computing the emitted power as
\begin{equation}
\label{dedt}
\frac{dE}{dt}=\frac{1}{32\pi}\sum_{\ell, m}
        \left(\left|\frac{d Q^{+}_{\ell m}}{dt}\right|^2+
        \left|Q^{\times}_{\ell m}\right|^2\right)\ ,
\end{equation}
the total energy lost to gravitational radiation is $E = 1.45 \times
10^{-6} (M/M_{\odot})$. This is about two orders of magnitude smaller
than the estimate made in~\cite{sp85} for a star with $a=0.54$, but
larger than the energy losses computed recently in the collapse of
rotating stellar cores to protoneutron stars~\cite{metal04}.

        In conclusion, we have presented the first waveforms from the
gravitational collapse of rapidly rotating stars to black holes using 3D
grids with Cartesian coordinates. The great potential shown by the PMR
techniques employed here opens the way to a number of applications that
would be otherwise intractable with uniform Cartesian grids. Work is in
progress to consider initial models with realistic EOSs or in
differential rotation, for which values $a \gtrsim 1$ can be reached and
more intense gravitational radiation is expected.

\bigskip
        We thank V.\ Ferrari, K.\ Kokkotas, N.\ Stergioulas, O.\ Zanotti
for discussions, and C.\ Ott and B.\ Zink for help with the PMR
algorithm. The computations were performed on {\em Albert100} at the
University of Parma and {\em Peyote} at the Albert Einstein Institute. ES
was supported by the DFG's SFB TR/7.

\vskip -0.25cm



\begin{thebibliography}{99}
\bibitem{note}{The frequencies for the QNMs of the black hole and for the
$w$-modes of the star are those most easily obtained from tabulated
values in the literature and serve here as indicative references.}
\end{thebibliography}
\end{document}